\documentclass[letter]{aa} 
\usepackage{graphicx}
\usepackage{txfonts}
\begin{document}
   \title{Evidence for an anticorrelation between the duration of the shallow decay phase of GRB X-ray afterglows and redshift}

   \author{G. Stratta
          \inst{1}
          \and
          D. Guetta\inst{2}
	  \and
	  V. D'Elia \inst{2}
	  \and
	  M. Perri \inst{1}
	  \and
	  S. Covino \inst{3}
	  \and
	  L. Stella \inst{2}
          }

   \institute{ASI Science Data Center, via G. Galilei, 00044 Frascati, Italy
             \thanks{INAF personnel resident at ASDC}\\
         \and
	INAF - Osservatorio Astronomico di Roma, via Frascati 33, 00040
		Monte Porzio Catone, Italy \\
 	 \and
	     INAF - Osservatorio Astronomico di Brera, via Bianchi 46, 23807, Merate (LC), Italy\\
             }

   \date{Received ....; accepted ....}

 \abstract{}{}{}{}{} 
 
  \abstract
   {One of the most intriguing features discovered by Swift is a plateau phase in the X-ray flux decay of about $70\%$ of the afterglows of gamma-ray bursts (GRBs). The physical origin of this feature is still being  debated. }
   {We constrain the proposed interpretations, based on the intrinsic temporal properties of the plateau phase.}
   {We selected and analyzed all the {\it Swift}/XRT GRB afterglows at known redshift observed between March 2005 and June 2008 featuring a shallow decay phase in their X-ray lightcurves.  }
   {For our sample of 21 GRBs we find an anticorrelation of the logarithm of the duration of the shallow phase with  redshift, with a Spearman rank-order correlation coefficient of $r=-0.4$ and a null hypothesis probability of $5\%$. When we correct the durations for cosmological dilation, the anticorrelation strenghtens, with $r=-0.6$ and a null hypothesis probability of $0.4\%$. Considering only those GRBs in our sample that have a well-measured burst peak energy (8 out of 21), we find an anticorrelation between the energy of the burst and the shallow phase duration, with r=-0.80 and a null hypothesis probability of $1.8\%$. }
{If the burst energy anticorrelation with the shallow phase duration is real, then the dependence of the shallow phase on redshift could be the result of a selection effect, since on average high-redshift bursts with lower energies and longer plateaus would be missed. A burst energy anticorrelation with the shallow phase duration would be expected if the end of the plateau arises from  a collimated outflow.  Alternative scenarios  are briefly discussed involving a possible cosmological evolution of the mechanism responsible for the X-ray shallow decay. 
}

   \maketitle
%

\section{Introduction}

In the pre-{\it Swift} era the X-ray afterglows of gamma-ray bursts could be observed only 
after many hours after the burst, when the flux typically 
showed a smooth power law-like decay, $t^{-\alpha}$, with an index of about 
$\alpha\sim 1$. Hereafter, we refer to this as the {\it standard} X-ray afterglow decay phase.
The {\it Swift} mission (Gehrels et al. 2004) has revolutionized GRB studies in many respects by observing the X-ray afterglow phase from a few dozen seconds after the burst (e.g. Zhang et al. 2007). 

The shallow decay phase observed in the X-ray flux of about $\sim 70\%$ of the afterglows (e.g. Panaitescu 2007) is one of the most intriguing features discovered by {\it Swift}. This phase usually becomes visible a few hundred seconds after the burst, after the steep decay in the prompt emission, and it lasts for $\sim1-10$ ks (Nousek et al. 2006; Zhang et al. 2006). No spectral evolution is observed in either the 0.3-10 keV range during the shallow phase or in the subsequent decay phase (e.g. Liang et al. 2007; Butler and Kochevski 2007). 
This lack of X-ray spectral variations has suggested that the observed X-ray temporal steepening is not associated with the crossing of a characteristic synchrotron frequency (e.g. the cooling frequency). 
Optical afterglow lightcurves often show a different behavior from those in the X-rays (Liang et al. 2007). This idicates either that the X-ray and optical afterglow have different origins or that the microphysical parameters determining the instantaneous energy in the electrons and magnetic field evolve in time (Panaitescu et al. 2006). Some of the models proposed to interpret the physical origin of the shallow decay phase are summarized in Zhang et al. (2007). 

Several studies have addressed the intrinsic properties of the X-ray shallow 
phase, in particular by testing whether a dependence exists between the 
intrinsic duration of the shallow phase and the burst energetics. Results obtained so far are 
discordant. An anticorrelation between the intrinsic duration of the shallow phase and the burst energetics has been found in some works (e.g. Sato et al. 2007; Dado et al. 2008), while in some others it was not (e.g. Liang et al. 2007; Nava et al. 2007). Sato et al. (2007) argue that these discrepancies may be associated with the large uncertainties affecting the burst energetics estimates and/or in modeling the X-ray shallow phase and estimating the temporal break between the shallow and the standard phases. Some of the discrepancies might also be ascribed to the different size and quality of the GRB samples used by different authors.

A linear dependence between the logarithm of the duration of the shallow phase, and the logarithm of the burst istropic equivalent energy would be expected if the temporal break between the shallow phase and the standard phase were interpreted as `jet break' time ($t_j$) and the GRB energy corrected for beaming factor were constant (e.g. Frail et al. 2001). Indeed, the jet opening angle can be estimated from the time $t_j$ at which the relativistic beaming ($1/\Gamma(t_j)$ where $\Gamma(t_j)$ is the fireball Lorentz factor) becomes equal to the geometric beaming of the fireball of half-opening angle $\theta_j$, that is, $\theta_j \propto (t_j^3/E_{iso})^{1/8}$ (Sari et al. 1999), where $E_{iso}$ is the equivalent isotropic energy.  At that time ($t_j$), the afterglow lightcurve decay steepens. The beamed corrected energy is $E_{\gamma}=E_{iso}(1-cos\theta_j)\sim E_{iso}\theta_j^2$. If $E_{\gamma}$ is constant, it follows that $E_{iso} \propto t_{j}^{-1}$. The correlation found by Ghirlanda et al. (2004) between the intrinsic peak energies and the beaming-corrected burst energies tells that the relation $E_{iso} \propto t_{j}^{-1}$ is still (nearly) valid for GRBs with similar intrinsic peak energy. In the jet scenario, the lightcurve steepening is expected to be achromatic. This condition is barely satisfied if we consider both the X-ray and the optical energy domains, since, as mentioned above, several X-ray shallow phases are not tracked in the optical regime. However, by restricting the energy range to the X-rays, the condition is satisfied since the lack of any spectral variation is a characteristic feature of the X-ray shallow phase and the subsequent standard decay phase.  

In the present paper we consider a sample of GRBs with a well-monitored X-ray lightcurve and known redshift, which unambigously showed a shallow decay phase. We find clear evidence of a redshift dependence of the duration of the shallow decay phase.

\begin{table*}
\centering
\begin{tabular}{|c|c|c|c|c|c|c|}
 \hline
GRB 		 & $ t_2$ [ks]  & $\alpha_1$ & $z$    &$E_{iso} [10^{52}]$erg  & $E_{peak,i}$ [keV]  &Comments	\\
\hline							       							       							       
\bf 050315 	& $9.0\pm0.2$   & $-0.2\pm0.1$  & 1.949  & $-$  	& $-$  & steep phase excluded\\
\bf 050319	& $2.7\pm0.1$   & $0.1\pm0.2$   & 3.240  & $-$	        & $-$   & $-$ \\
050401 		& $5.0\pm0.3$   & $0.63 \pm0.02$& 2.90   & $35\pm7^A$ 	 & $467\pm110^{(1)}$ &$-$ \\
050505          & $6.4\pm0.4$   & $0.1\pm 0.1$  & 4.27   & $19.5\pm3.1^N$& $622\pm211^{(2)}$ &$-$ \\
051109A 	& $2.0\pm0.1$   & $0.54\pm0.03$ & 2.346  & $6.5\pm0.7^A$ & $539\pm200^{(1)}$ & $-$ \\
060502A     	& $30\pm2$      & $0.43\pm 0.09$& 1.51   & $-$	         & $-$          &$-$ \\
060526 	        & $20\pm2$ 	& $0.11\pm 0.07$& 3.21   & $2.6\pm0.3^A$ & $105\pm21^{(1)}$ & flare excluded\\
060607A 	& $12.9\pm0.2$  & $0.44\pm 0.02$& 3.082  & $12.2\pm1.8^N$& $535\pm164^{(2)}$ &flare excluded\\
\bf 060614 	& $47\pm2$      & $0.05\pm 0.03$& 0.125  & $0.21\pm0.09^A$&$55\pm45^{(1)}$&steep phase excluded\\
\bf 060714 	& $3.6^{+1.2}_{-0.7}$ & $0.24\pm 0.05$& $2.71$ & $-$  &$-$  &flare excluded\\
\bf 060729 	& $77\pm1$      & $0.12\pm 0.02$& 0.54   &  $-$ 	&$-$  & flare excluded\\
\bf 060814 	& $9.9\pm0.2$   & $0.25\pm0.06$ & 0.84   & $7.0\pm0.7^A$& $473\pm155^{(1)}$& flare excluded\\
\bf 061121 	& $3.6\pm0.5$   & $0.25\pm0.04$ & 1.314  & $22.5\pm2.6^A$& $1289\pm153^{(1)}$&$-$ \\
\bf 070306 	& $27.2\pm0.8$  & $0.11\pm0.02 $& 1.4959 & $-$ 	& &$-$ \\
070529          & $2^{+2}_{-1}$   & $0.7\pm0.1$ & 2.4996 & $-$	& $-$ & $-$ \\
070611          & $50^{+10}_{-13}$& $0.1\pm0.1$ & 2.04   & $-$ 	& $-$ &$-$ \\
070802          & $5^{+2}_{-1}$   & $0.1\pm0.2$ & 2.45   & $-$	& $-$ &$-$ \\
080310          & $4.5\pm0.9$   & $0.7\pm0.1$   & 2.42   & $-$ 	& $-$ &flare excluded\\
\bf 080430      & $15^{+8}_{-3}$   & $0.4\pm0.1$& 0.767  & $-$	& $-$ &$-$ \\
080605          & $0.55\pm0.03$  & $0.68\pm0.04$& 1.6398 & $-$ 	& $-$  &$-$ \\
080607          & $1.5\pm0.2$ & $0.1^{+0.9}_{-0.3}$& 3.036& $-$ & $-$  &flare excluded\\
\hline
\end{tabular}
\caption{\label{sample}
Selected sample of GRBs with known redshift that presents a shallow decay phase.
}
$^{(1)}$ From Amati et al. 2008;
$^{(2)}$ From Nava et al. 2007.
\end{table*}

  \begin{figure}
   \centering
	\includegraphics[width=7cm]{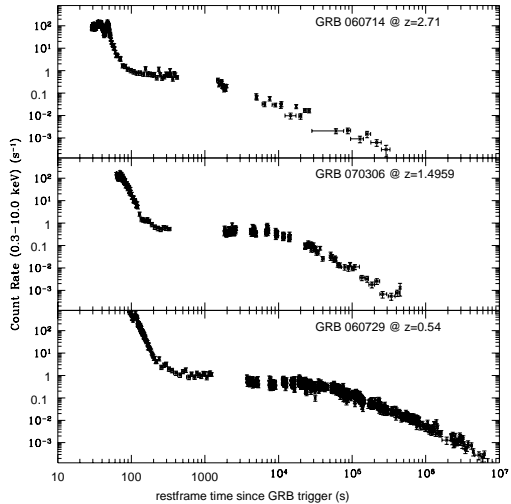}
   \caption{X-ray afterglow lightcurves from Evans et al. (2007) where the temporal axis has been  rescaled to the burst rest frame for 3 GRBs of our `golden sample' a different redshift (see \S 2). }
              \label{f:f1}%
    \end{figure}


\section{The sample and data analysis}

The X-ray afterglow lightcurves supplied by the UK Swift Science Data Centre at the University of Leicester were used (Evans et al. 2007). The sample was built by selecting in the Swift archive all GRBs observed in the period from March 2005 to June 2008 with the following characteristics: i) a 0.3-10 keV XRT lightcurve featuring shallow behavior with temporal index $\alpha_1<0.8$ (i.e. shallower than the `standard' fireball model predictions, e.g. Sari et al. 1998), over a temporal interval greater than $0.5$ ks (so that the power-law decay index can be measured accurately); ii) the shallow decay interval should not be dominated by features such as spikes or flares that may affect the measurement of the decay index; iii) since our analysis concentrates on the study of the intrinsic properties of the shallow phase, we considered only GRBs at known distances. In this way we selected 21 GRBs (out of a total of about 60 bursts at known redshift for which the statistics were good enough to carry out a detailed temporal analysis). The sample, as well the redshift of each burst, is given in Table 1. In Figure \ref{f:f1} the lightcurves of three GRBs from our `golden sample' (see below), taken from Evans et al. (2007), are plotted after rescaling of the temporal axis to their rest frame. The shape of the lightcurves clearly shows the well known `canonical' behavior (e.g. Nousek et al. 2005). Since our goal is to measure the duration of the shallow phase, we fitted the observed lightcurves with power laws as: 
\begin{equation}
        \label{eq:brapp}
        F_\nu(t) \propto
        \left\{
        \begin{array}{ccc}
                t^{-\alpha_0} & t< t_1 \\
                t^{-\alpha_1}   &  t_1<t<t_2 \\
		t^{-\alpha_2} &  t>t_2
        \end{array}
        \right.
\end{equation}

In Eq. (1), $t<t_1$ indicates the interval where the typical initial steep decay from the prompt is observed, $t_1<t<t_2$ corresponds to the shallow decay phase, and $t>t_2$ to the subsequent standard decay. In this work, we consider $t_2$ as a measure of the duration of the shallow decay. For GRB 060614 and GRB 060814, we excluded the initial steep decay from the fit because it could not be well-fitted by a power-law decay. We also excluded from the fit those intervals in which flares were present for $t<t_1$ in the lightcurves of other GRBs (see last column of Tab. 1). 

Table \ref{sample} shows the selected dataset where $t_2$ represents the observed epoch at which the shallow phase steepens to the standard decay (calculated from the burst onset as determined by Swift/BAT), and $\alpha_1$ represents the temporal index of the shallow decay region. Errors are given at the $1\sigma$ confidence level. 
In some cases Eq. (1) provided a poor approximation of the steepening from the shallow phase to the standard one. We thus checked whether other estimates of the temporal break between the shallow phase and the standard phase obtained assuming more complex models (e.g. Willingale et al. 2007; Ghisellini et al. 2008) provided different results and we find no significant differences within the uncertainties, except for two cases\footnote{Willingale et al. (2007) estimated for GRB 060607A a duration of $T_a=56^{+4}_{-3}$ ks and Ghisellini et al. (2008) for GRB 050319 estimated $T_A=7$ ks (the latter is in the rest frame).}.

We first checked whether any common intrinsic value $t^{'}_{2}$ of the epoch at which the X-ray shallow decay ends ($t^{'}_{2}=t_2/(1+z)$) exists for all GRBs. The observed epoch $t_2$ covers 3 orders of magnitudes (0.5-80 ks).  The intrinsic epoch $t^{'}_{2}$ still covers a wide range of values (0.2-50 ks, Tab. 1). We find no evidence of clustering around any particular value. We then checked whether there is any redshift dependence on $t_2$. We found that an  anticorrelation exists between the logarithm of $t_2$ and $z$, with a rank correlation factor of -0.4. With 19 degrees of freedom, the null hypothesis is rejected at a $95\%$ confidence level. After correcting $t_2$ for cosmological dilation, we found that the anticorrelation is strengthened, with a rank correlation factor of -0.6: the null hypothesis is now rejected at $99.6\%$ (non-directional probability) confidence level (Fig. \ref{f:f2}). That the anticorrelation already present for $t_2$ becomes more significant after correction for cosmological dilation provides evidence that the correlation is genuine and not biased by the redshift correction. To confirm this result, we selected a ''golden sample" of 10 GRBs from our original 21 GRB sample by considering only those GRBs with the best XRT coverage in all the three typical {\it Swift}/XRT X-ray lightcurve components (e.g. Nousek et al. 2005), which are an initial steep decay followed by the shallow and then standard decay. Despite the decrease in the sample of GRBs, the anticorrelation between the logarithm of $t_2$ and $z$ persists, with a rank correlation factor of -0.85. (The null hypothesis is rejected at $99.2\%$ confidence level.) The golden sample is marked with red open circles in Figure \ref{f:f2} and in boldface in Table 1.

These findings cannot be interpreted as due to an energy dependence of the duration of the shallow phase as one may conclude for example from Figure \ref{f:f1} where the shallow phase at energies of 1-37 keV (for GRB 060714 at z=2.711) in the rest frame is shorter than the one at 0.5-15 keV (for GRB 060729 at z=0.54). In fact, it is well known that the hardness ratio, defined as the flux ratio in the 0.3--1.5 keV and 1.5--10 keV energy bands, does not show any evidence of variations along the shallow phase (e.g. Butler and Kocevski 2007; Liang et al. 2007).  

The anticorrelation of the intrinsic duration of the X-ray shallow phase with redshift that we discussed above may be a consequence of an anticorrelation of $t^{'}_{2}$ with burst energy. Indeed, the GRBs that we observe at high-redshift are on average more energetic than the low-redshift ones due to a simple selection effect. We cannot see faint GRBs at large distances. Therefore, shorter plateaus might be observed more frequently at high redshift because associated with more energetic GRBs. Alternatively, the anticorrelation of $t^{'}_{2}$ with $z$ can be explained as a cosmological evolution of the mechanism that gives rise to the shallow decay. We briefly discuss both possibilities here.

\subsection{The burst energy dependence of $t^{'}_{2}$}

\begin{figure}
   \centering
   \includegraphics[width=9cm]{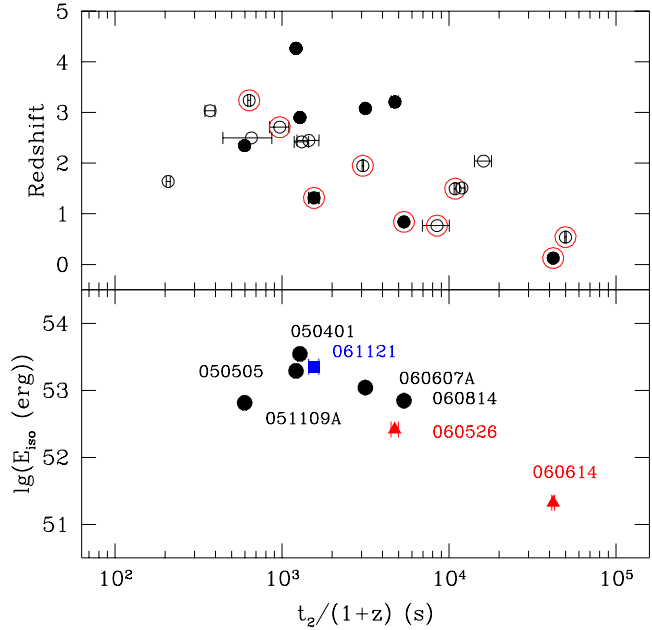}
   \caption{{\it Top panel}: redshift versus the intrinsic duration of the shallow phase. The observed anticorrelation has a null hypothesis rejected at $\ge99.6\%$ confidence level. The filled black circles are the GRBs with measured peak energy and those encircled in red represent the `golden sample' (in bold face in Tab. 1). The rest of the sample is plotted with open circles. {\it Bottom panel}: 
the isotropic equivalent energy of the burst versus the intrinsic duration of the shallow phase. The observed anticorrelation has a null hypothesis rejected at $\ge98.2\%$ confidence level. Black circles, red triangles, and blue square indicate GRBs with similar intrinsic peak energy ($400 \le E_{peak,i}\le 600$ keV,  $E_{peak,i} \le 200$ keV and $E_{peak,i} \ge 1000$ keV respectively, see Table 1. 
}
              \label{f:f2}%
    \end{figure}

To verify the dependence of the energy of the burst from $t^{'}_{2}$ is not an easy goal since the GRB energetics, usually estimated by $E_{iso}$, is often affected by large uncertainties in the burst spectral parameters, the peak energy $E_{peak}$ of the $EF_E$ spectrum in particular. For this reason, we considered only those GRBs of our sample for which precise measurements of $E_{peak}$ are available (e.g. Amati et al. 2008; Nava et al. 2007). For 8 GRBs of our sample (3 of which are part of the `golden sample', see Tab. 1 and Figure \ref{f:f2}), we find that $t^{'}_{2}$ anticorrelates with $E_{iso}$, with a rank correlation factor of --0.80 and null hypothesis rejected at $98.2\%$ confidence level (Fig. \ref{f:f2}). If we restrict ourselves to considering only those GRBs with similar intrinsic peak energy, we still find marginal evidence of the anticorrelation, although the statistics are poor (Fig. \ref{f:f2}).
Even though a firm conclusion could not be reached, this result is consistent with the results obtained by Dado et al. (2008) and Sato et al. (2007), while at odds with findings by Liang et al. (2007) and Nava et al. (2007). 

As mentioned in \S1, the anticorrelation between $t^{'}_{2}$ and $E_{iso}$ suggests that $t_2$ could be considered as the jet break time of (the part of) the fireball that gives rise to the X-ray afterglow (see \S 1). 
Several suggestions have been made to reconcile the jet interpretation with features observed in some GRB, such as a chromatic evolution of the break or the temporal break at later epochs. For example, the two-component jet model (e.g. Peng et al. 2005; Racusin et al. 2007) explains the lack of a simultaneous optical break as a deficit in the optical emission from the narrower of the two jets,  responsible for the two X-ray temporal breaks. In another model, Ghisellini et al. (2007) propose to interpret the observed X-ray plateau as the sum of two components: the late prompt emission (internal shock between late emitted shells) and the afterglow. The temporal break at the end of the X-ray shallow decay phase is the proof that the Lorentz factor of the late shells (typically smaller than that of the external shell giving rise to the afterglow) has reached the $1/\theta_j$ value (see \S1).  Whether the break is tracked in the optical band depends on the relative intensity of each component. The second break at later times is produced when the Lorentz factor of the shell producing the afterglow has reached the $1/\theta_j$ value.

\subsection{Evidence of cosmic evolution?}

Given the uncertainties affecting the burst energetic evaluations and the poor statistics available to definitively confirm the anticorrelation between $E_{iso}$ and $t^{'}_{2}$, we 
speculate on another possible interpretation of the redshift anticorrelation with $t^{'}_{2}$ where the shallow phase depends on an external component that evolves with redshift. This is the case, for example, if the shallow phase is produced by the interaction of the X-ray emission with the surrounding interstellar dust (e.g. Klose 1998,1999; Shao and Dai 2007,2008). 
The expected decrease in the interstellar dust content with redshift might be the reason for the observed anticorrelation between the duration of the shallow phase and redshift. Moreover, it is expected within this model that the X-ray and optical temporal breaks are in general uncorrelated, in agreement with a number well-sampled afterglow lightcurves (Liang et al. 2007). However, this interpretation faces problems in explaining the lack of the predicted spectral variation in the X-ray spectra, as already pointed out by Shen et al. (2008). A possible solution might involve complex dust distribution along the line of sight.
Other scenarios are still possible, as for example if the X-ray shallow phase depends on the intrinsic GRBs properties (e.g. inner engine) that may evolve within cosmological time scales. 

\section{Conclusions}

In this work we analyzed 21 GRBs with known redshift that feature a shallow phase in the X-ray lightcurve. Our main result is a clear anticorrelation of the intrinsic duration of the X-ray shallow phase with redshift. Considering only those GRBs in our sample that have well-measured burst peak energy, we find marginal evidence for burst energy anticorrelation with the shallow phase duration.  
The latter anticorrelation would be expected if the observed temporal break ($t_2$) arises from  a collimated outflow. In this case, the $t^{'}_{2}$ anticorrelation with $z$ can be interpreted as the evidence of a selection effect since high-redshift bursts with lower energies and shorter  plateaus would be missed.   However, a larger sample of bursts at known redshift with well-measured burst spectral parameters is required to definitively assess the $E_{iso}$ anticorrelation with $t^{'}_{2}$.
In an alternative scenario, the shallow phase may arise from a mechanism that operates differently at high redshift, such as for example from X-ray dust scattering or an evolution of the intrinsic GRB properties as for example the inner engine (Guetta et al. in preparation). 
Finally we note that by virtue of its redshift dependence (though with large scatter), the {\it observed} X-ray shallow phase duration ($t_2$) may be regarded as an additional figure of merit to single out high-redshift GRBs directly from X-rays observations. This might provide useful information for burst follow-up campaigns at optical and NIR wavelengths.

\begin{acknowledgements}
We thank the anonymous referee for his/her useful comments. This work is supported in Italy from ASI Science Data Center and by ASI grant I/024/05/0 and MIUR grant 2005025417.
\end{acknowledgements}

\end{document}